\documentclass[longbibliography,noeprint,noshowpacs,nopreprintnumbers,twocolumn,pra,showpacs,superscriptaddress,amsmath,amssymb,citeautoscript,aps,10pt]{revtex4-2}
\usepackage{graphicx}
\usepackage{dcolumn}
\usepackage{color}
\usepackage{booktabs}
\usepackage{bm}
\usepackage[hidelinks]{hyperref}
\hypersetup{
    colorlinks,
    citecolor=blue,
    filecolor=blue,
    linkcolor=blue,
    urlcolor=blue
}

\DeclareMathOperator{\sgn}{sgn}

\graphicspath{{Figures/}{./Figures/}{.}}

\allowdisplaybreaks

\begin{document}

\title{Torsion Induced Asymmetric Luttinger Liquids}
\author{Arseny Pantsialei}
\affiliation{Institute of Physics, Maria Curie-Sk\l{}odowska University, 20-031 Lublin, Poland}
\author{Nicholas Sedlmayr}
\email[e-mail: ]{sedlmayr@umcs.pl}
\affiliation{Institute of Physics, Maria Curie-Sk\l{}odowska University, 20-031 Lublin, Poland}

\date{\today}

\begin{abstract}
We consider a general model of a Luttinger liquid with broken parity and time reversal symmetry, but with their composite symmetry intact. Such a scenario can be due to a combination of torsion and a Zeeman field in nanowires, or a result of bringing different helical Luttinger liquids into proximity. The broken symmetries result in a band structure with no axis of symmetry, and therefore with asymmetric velocities between left and right moving contributions. By taking a general spin-full model with all possible scattering and interaction terms in the bosonic model we show that generically the spin degree of freedom becomes gapped out, resulting in an effective spinless Luttinger liquid with asymmetric velocities. Our work generalizes and extends previous studies which focused on a minimal model of a spinless Luttinger liquid. We further demonstrate that a possible experimental signature of the asymmetry of such asymmetric models can be seen in the spectral function.
\end{abstract}  

\maketitle

\section{Introduction}\label{sec_intro}

It is well known that Landau's Fermi liquid theory breaks down for electrons in a one dimensional wire due to the importance of backscattering in this case~\cite{Voit1995,Maslov2005}. The Landau Fermi liquid is replaced instead by a Luttinger liquid~\cite{Tomonaga1950,Luttinger1963,Giamarchi2004} which describes a universality class of gapless one dimensional quantum systems~\cite{Haldane1981,Haldane1981a}. Included in this class are some interacting fermionic and bosonic systems, as well as some spin chains. A characteristic feature is the separation of spin and charge degrees of freedom in a Luttinger liquid~\cite{Auslaender2005,Yao1999,Bockrath1999,Jompol2009,Deshpande2010}. Another is the tendency for local impurities to lead to effectively broken wires at low energies~\cite{Kane1992,Kane1992a,Eggert1992,Furusaki1993,Furusaki1996,Pereira2004,Sedlmayr2011b}. Experimental signatures of a Luttinger liquid can most easily be seen in measurements of conductance scaling as a function of parameters such as the filling fraction or temperature~\cite{Liang2001,Javey2003,Yacoby1996,Steinberg2008,Tarucha1995,Purewal2007,Kamata2014}, or in the behavior and scaling of the density of states~\cite{Giamarchi2004}.

The standard procedure for deriving the Luttinger liquid for an interacting one dimensional system of fermions starts by linearizing the non-interacting band structure. Interactions are included via bosonization. The nature of the bands at the Fermi level can therefore play a profound role in the nature of the final Luttinger liquid~\cite{Fernandez2002,Sedlmayr2013b}. Regarding fermionic models there are many generalizations of the basic Luttinger theory, extending it to Helical Luttinger liquids~\cite{Braunecker2012,Meng2014,Calzona2015,Li2015,Braunecker2018,Stuhler2019,Hsieh2020,Zakharov2024}, or spin-orbit coupling in carbon nanotubes~\cite{Schulz2009,Schulz2010} and in magnetized spin chains or quantum wires~\cite{Gangadharaiah2008,Sedlmayr2013b}. Inhomogeneity in the one dimensional system, particularly due to contacting the wire to leads for the purpose of calculating transport, has also received wide attention~\cite{Yue1994,Safi1995,Maslov1995,Ogata1994,Wong1994,deC.Chamon1997,Safi1999,Imura2002,Enss2005,Janzen2006,Rech2008,Rech2008a,Gutman2010,Thomale2011,Sedlmayr2012a,Sedlmayr2013a,Sedlmayr2014,Morath2016}. Most significantly it is found that the expected rescaling of conductance by the Luttinger parameter is not visible in actual measurements of the conductance~\cite{Safi1995,Maslov1995}.

Another form of helicity seen in wires is that caused by torsion~\cite{Ortix2015,Gentile2015}. In this case the electronic states gain a handedness caused by the winding of the wire in space, \emph{i.e.~}a torsion, which manifests as a spin-orbit like coupling in the Hamiltonian~\cite{Ortix2015}. As we will show this can lead to an asymmetry in left/right moving velocities of the bands at the Fermi level when a magnetic field is applied, which importantly can not be removed by a boost in reciprocal space. These considerations motivate us to introduce a very general linearized model of spinfull fermions with broken parity and time-reversal symmetry. The composite symmetry is assumed to still hold. We allow all scattering and interaction terms in our model consistent with these symmetry requirements. Previously only a minimal spinless model with an asymmetric velocity term and the simplest density-density interaction has been considered~\cite{Fernandez2002}. We show that generically the low energy theory has a gapped spin degree of freedom, and diagonal chiral bosonic charge modes with asymmetric velocities. We propose the spectral function as an experimental signature of the asymmetry. We note that, as discussed in \onlinecite{Fernandez2002}, asymmetric bands can also be obtained by bringing into proximity chiral edge modes of different Quantum Hall systems.

This article is organized as follows. We begin in Sec.~\ref{sec_micro} by considering microscopic mechanisms for the required band asymmetry. Then in Sec.~\ref{sec_wires} we introduce our linearized model, discuss its symmetry properties and bosonize it. Sec.~\ref{sec_low} contains a renormalization group analysis of the massive spin term in the Hamiltonian and derives the effective low energy single mode model.  In Sec.~\ref{sec_spec} we calculate the spectral function and show it is a possible experimental test of the asymmetry and in Sec.~\ref{sec_con} we conclude. Throughout this article we use units where $\hbar=1$ and $k_B=1$.

\section{Microscopic and Geometric Origins of the Tilt}\label{sec_micro}

Before introducing the general interacting model, we give a possible mechanism by which the velocity asymmetry can be created. We demonstrate that for a quasi-one-dimensional channel projected onto the two low-energy bands, the combined presence of torsion $\tau$ and a Zeeman field $B$ generates a genuine odd-in-$k$ contribution to the projected dispersion. This contribution is not equivalent to a simple momentum shift of a symmetric bandstructure and therefore produces a finite spectral asymmetry, a ``tilt''. 

One can quickly show that if the bandstructure has an asymmetry which can be removed by a boost $k\mapsto k-k_0$ then the Fermi velocities will have the same magnitude. Taking a single spinless band which we assume \emph{can} be written as a boosted even in momentum dispersion,
\begin{equation}
\epsilon(k)=\tilde\epsilon(k-k_0)\textrm{ such that } \tilde\epsilon(q)=\tilde\epsilon(-q)\,,
\label{boost_form}
\end{equation}
the group velocities at the Fermi points are
\begin{equation}
v_{\pm}=\pm\partial_k\epsilon(k_{F\pm})\,.
\label{u_def}
\end{equation}
Note that the velocities $v_{\pm}$ are defined here to be the magnitudes, \emph{i.e.}~positive. The Fermi points are $k_{F\pm}=k_0\pm k_F$, where $\tilde\epsilon(\pm k_F)=\epsilon_F$. Since $\partial_q\tilde\epsilon(q)$ is odd, we have
$\partial_k\epsilon'(k_{F-})=\partial_k\tilde\epsilon(-k_F)=-\partial_k\tilde\epsilon(k_F)=-\partial_k\epsilon(k_{F+})$, and \eqref{u_def} immediately gives $v_+=v_-$. This merely amounts to saying that a pure boost can shift the Fermi points, but it cannot make the two Fermi-point slopes unequal.

A non-zero tilt therefore requires a dispersion with no axis of symmetry. A minimal microscopic realization is provided by a spinful nanowire on a helical space curve with spin-orbit coupling and Zeeman splitting. Following Ref.~\onlinecite{Ortix2015} we consider a one-dimensional electronic system with spin confined to a nanowire. The axis of the wire is a space curve 
$\mathbf{r}(s)$ parameterized by the arc length $s$. The local geometry of the curve is described by an orthonormal moving frame $(\{\hat{\mathbf{t}}(s), \hat{\mathbf{n}}(s), \hat{\mathbf{b}}(s)\}$; the tangent, normal, and binormal vectors;
with curvature $\kappa(s)$ and torsion $\tau(s)$:
\begin{equation}\label{frenesere}
\partial_s \hat{\mathbf{t}}=\kappa \hat{\mathbf{n}}\,,\quad
\partial_s \hat{\mathbf{n}}=-\kappa \hat{\mathbf{t}}+\tau \hat{\mathbf{b}}\,,\textrm{ and }
\partial_s \hat{\mathbf{b}}=-\tau \hat{\mathbf{n}}\,.
\end{equation}
We consider the single-channel limit, in which only the lowest transverse mode is occupied, so that the low-energy dynamics becomes effectively one-dimensional along the arc coordinate $s$. The geometry of the wire is assumed to be smooth: the curvature $\kappa(s)$ and the torsion $\tau(s)$ vary slowly on the scale of the Fermi wavelength. We retain only the leading term that is linear in the torsion and linear in the Zeeman energy, and neglect higher-order corrections.

With these assumptions the reduction of the curved-wire problem yields an effective one-dimensional Hamiltonian for a fermion of mass $m$:
\begin{equation}
H_{1\mathrm D}=\frac{1}{2m}\bigl(p_s-A_s(s)\bigr)^2-\mu + \Phi_s(s)
+\frac{1}{2} \mathbf B\cdot \boldsymbol{\sigma},
\label{H1D_su2}
\end{equation}
where $p_s=-i\partial_s$, $A_s(s)$ is a non-abelian $SU(2)$ gauge field generated by the spin-orbit coupling and the moving frame. $\Phi_s$ collects all geometric scalar terms from the reduction, $\mu$ is the chemical potential, and the last term is a Zeeman field $\mathbf{B}$.

The gauge field is
\begin{equation}
A_s(s)=\mathbf a(s)\cdot\boldsymbol{\sigma}.
\label{As_asigma}
\end{equation}
For a uniform segment of the wire, where the curvature $\kappa$ and torsion $\tau$ are locally constant, the vector $\mathbf a$ has 
the general structure
\begin{equation}
\mathbf a = \alpha m\hat{\mathbf u}
+\frac{\kappa \hat{\mathbf b}}{2}
+\frac{\tau \hat{\mathbf t}}{2}
+\ldots ,
\label{a_structure}
\end{equation}
The first term comes from the spin-orbit coupling (the unit vector $\hat{\mathbf u}$ is determined by the microscopic geometry of the spin-orbit interaction), while the terms proportional to $\kappa$ and $\tau$ arise from the rotation of the moving frame along the curve. A local rotation of the moving frame corresponds to a local spin rotation $U(s)\in \mathrm{SU}(2)$. Under such a transformation the 
connection changes according to
\begin{equation}
A_s  \to  U A_s U^\dagger - iU\partial_s U^\dagger .
\label{SU2_gauge}
\end{equation}
The gauge degree of freedom here is the choice of the reference frame.

\begin{figure}[t]
    \centering
    \includegraphics[width=0.8\columnwidth]{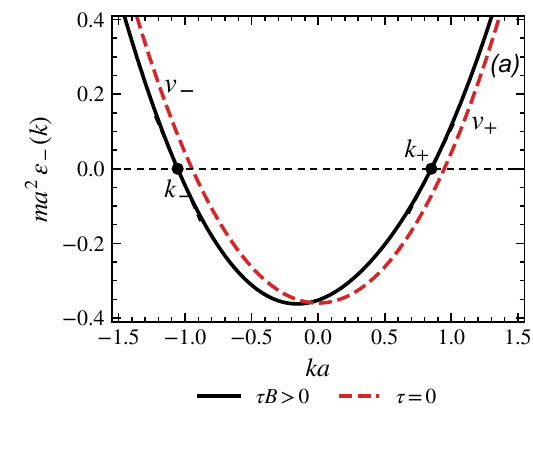}\\
    \includegraphics[width=0.8\columnwidth]{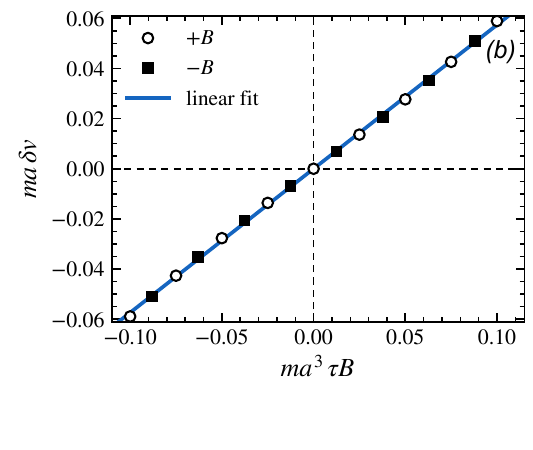}
    \caption{Microscopic extraction of the tilt from the exact projected lower band. In panel (a) the exact lower-band dispersion $\epsilon_{-}(k)$ for $\tau B>0$, $\tau B<0$, and $\tau=0$, are shown. For the representative asymmetric case, the tangents at the two Fermi points $k_-$ and $k_+$ are shown explicitly, demonstrating unequal chiral slopes $v_- \neq v_+$. Reversing the sign of $\tau B$ reverses the asymmetry. In panel (b) the velocity asymmetry or tilt, $\delta v=(v_--v_+)/2$, extracted numerically from the exact Fermi-point slopes is plotted versus $\tau B$. The odd dependence and near-linear behavior close to the origin confirm the microscopic scaling $\delta v\propto \tau B$.}
    \label{fig:micro_tilt}
\end{figure}

The full projected dispersion relation is
\begin{equation}
\epsilon_\mp(k)=\epsilon_0(k)\mp
\sqrt{\frac{k^2}{m^2}|\mathbf{a}|^2+
\frac{B^2}{4}-\frac{k}{m}(\mathbf{a}\cdot\mathbf{B})}
\end{equation}
where
\begin{equation}
\mathbf{a}=\mathbf{a}_0+\frac{\tau \hat{\mathbf{t}}}{2}\,.
\end{equation}
Expanding to leading order gives
\begin{equation}
\epsilon_-(k)\approx\epsilon_0(k)-D_0(k)
+\frac{\tau}{4m}\frac{k}{D_0(k)}\hat{\mathbf{t}}\cdot \hat{\mathbf{B}}
\end{equation}
and
\begin{equation}
D_0(k)=
\sqrt{\frac{k^2}{m^2} |\mathbf{a}_0|^2+\frac{B^2}{4}}\,.
\end{equation}
For uniform $\kappa,\tau$ and fields one finds an explicit odd-in-$k$ term
\begin{equation}
\delta\epsilon_{\mathrm{odd}}(k)
=
\frac{\tau}{4m} 
\frac{k}{D_0(k)}\hat{\mathbf t}\!\cdot\!\mathbf{B}\,,
\end{equation}
where $D_0(k)$ is an even function of $k$ (set by spin-orbit and Zeeman scales). This directly yields a finite tilt,
\begin{equation}
 \delta v=\frac{v_--v_+}{2}\propto \tau B\,.
\end{equation}
In the following we are interested only in the linearized form. Relevant parameters can be extracted from the form of the projected bandstructure, see Fig.~\ref{fig:micro_tilt}.

\section{General Interacting Asymmetric Wires}\label{sec_wires}

We start now from a model of spinful fermions with linearized dispersion relations at the Fermi energy. A general free linearized Hamiltonian for electrons with spin $\sigma$ reads
\begin{equation}
    H_{0f}=-i\sum_{\substack{\alpha\in\{\pm1\}\\\sigma\in\{\uparrow,\downarrow\}}}\alpha v_{\alpha\sigma}\int dx
    \psi^\dagger_{\alpha\sigma}\partial_x\psi_{\alpha\sigma}\,,
\end{equation}
with the linearized field operators given by
\begin{equation}
    \psi_\sigma(x)=\sum_{\alpha\in\{\pm1\}}e^{i\alpha k_{F\alpha\sigma}x}\psi_{\alpha\sigma}(x)\,.
\end{equation}
Here we have allowed arbitrary Fermi momenta $k_{F\alpha\sigma}$ and velocities $v_{\alpha\sigma}$, shortly we will impose the symmetry restrictions relevant for our problem. $\alpha$ labels the positive and negative Fermi momenta, and we will assume that the band velocities are positive/negative at the positive/negative Fermi momenta. For simplicity we also take $k_{F\alpha\sigma}=k_{F\sigma}$.

Time reversal symmetry $T=-i\sigma^y K$, with $K$ charge conjugation, would require
\begin{equation}
    v_{\alpha\sigma}=v_{\bar\alpha\bar\sigma}\,.
\end{equation}
However, this symmetry is broken by the magnetic field. Similarly, in the absence of torsion we have parity, which requires
\begin{equation}
    v_{\alpha\sigma}=v_{\bar\alpha\sigma}\,.
\end{equation}
If we require only $PT$ symmetry, but individually they are broken, then this constrains
\begin{equation}
    v_{\alpha\sigma}=v_{\alpha\bar\sigma}=v_\alpha\,.
\end{equation}
\emph{I.e.}~the two right moving modes must have the same velocity, but there is no restriction on having different left and right moving velocities. Our general non-interacting model thus becomes
\begin{equation}
    H_{0f}=-i\int dx \sum_{\alpha,\sigma}\alpha v_{\alpha}
    \psi^\dagger_{\alpha\sigma}\partial_x\psi_{\alpha\sigma}\,.
\end{equation}

\begin{figure}
    \centering
\includegraphics[width=0.95\columnwidth]{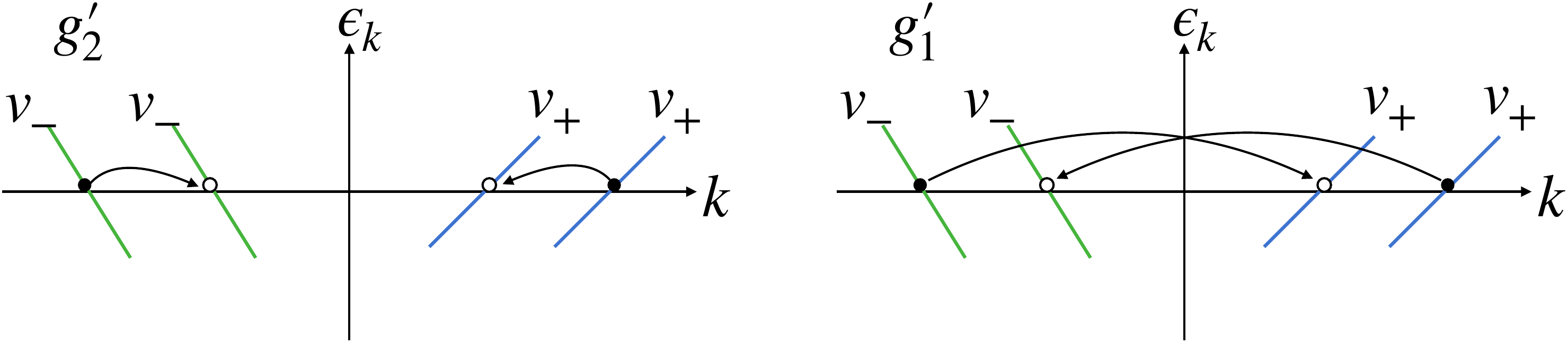}
\caption{Schematics of the $g'_2$ and $g'_1$ backscattering processes, as labeled on the figures. Pictured are those non-oscillating processes which lead to the massive term Eq.~\eqref{massive}.}\label{fig_schem}
\end{figure}

We also assume the presence of a screened density-density like interaction. We will impose $PT$ symmetry also on the interaction terms, but make no other restrictions. For density-density like terms we find, with $\rho_{\alpha\sigma}=\psi^\dagger_{\alpha\sigma}\psi_{\alpha\sigma}$ and the usual g-ology:
\begin{align}
    H_2&=\int dx\sum_{\alpha\sigma}\left[\frac{g_{2\parallel}}{2}\rho_{\alpha\sigma}\rho_{\bar\alpha\sigma}+\frac{g_{2\perp}}{2}\rho_{\alpha\bar\sigma}\rho_{\bar\alpha\sigma}\right]\,,\\
    H_4&=\int dx\sum_{\alpha\sigma\sigma'}\frac{g_{4\alpha}}{2}\rho_{\alpha\sigma}\rho_{\alpha\sigma'}\,,\textrm{ and}\\
    H_{1\parallel}&=-\int dx\sum_{\alpha\sigma\sigma'}\frac{g_{1\parallel}}{2}\rho_{\alpha\sigma}\rho_{\bar\alpha\sigma}.
\end{align}
$g_{1\parallel}$ is found from rearranging a backscattering term, and will be assumed in the following to have already been rescaled into $g_{2\parallel}$ (as will other equivalent terms). The bandstructure also allows for several backscattering terms under $PT$ symmetry. Assuming incommensurate filling so all oscillating terms vanish, we have
\begin{equation}
    H_{2'}=\sum_{\alpha\sigma}\int dx\frac{g_{2}'}{2}\psi^\dagger_{\alpha\sigma}\psi_{\alpha\bar\sigma}\psi^\dagger_{\bar\alpha\sigma}\psi_{\bar\alpha\bar\sigma}\,,
\end{equation}
and
\begin{equation}
    H_{1'}=\sum_{\alpha\sigma}\int dx\frac{g_{1}'}{2}\psi^\dagger_{\alpha\bar\sigma}\psi_{\bar\alpha\sigma}\psi^\dagger_{\bar\alpha\bar\sigma}\psi_{\alpha\sigma}\,.
\end{equation}
The $g'_{1}$ term has the same operator content as $g'_{2}$ term and we will assume it has been rescaled into that term. This is clear when looking at the scattering diagrams given in Fig.~\ref{fig_schem}.

We now bosonize the model with the vertex operator
\begin{equation}\label{vertex}
    \psi_{\alpha\sigma}(x)=\frac{1}{\sqrt{2\pi a}}e^{i\alpha\sqrt{2\pi}\phi_{\alpha\sigma}(x)}
\end{equation}
where $a$ is the lattice spacing, which will operate as a small distance cut-off. The field $\phi(x)$ obeys the commutation relation
\begin{equation}
    \left[\phi_{\alpha\sigma}(x),\phi_{\alpha'\sigma'}(x')\right]=\delta_{\alpha\alpha'}\delta_{\sigma\sigma'}\frac{i\alpha}{2}\sgn(x-x')\,.
\end{equation}
One can also show that $\sqrt{2\pi}\rho_{\alpha\sigma}=-\partial_x\phi_{\alpha\sigma}$, together these expressions allow us to find a bosonic form for the full Hamiltonian. More details of this procedure can be found in appendix \ref{app_details}. For $H_{0f}$ we find
\begin{equation}
    H_{0f}=\int dx\sum_{\alpha\sigma}\frac{v_{\alpha}}{2}\left[\partial_x\phi_{\alpha\sigma}(x)\right]^2.
\end{equation}
Collecting all quadratic terms, including those from the density-density interactions, and defining
\begin{equation}
    \Phi(x)=(\phi_{+1,\uparrow},\phi_{+1,\downarrow},\phi_{-1,\uparrow},\phi_{-1,\downarrow})^T
\end{equation}
we have
\begin{equation}
    H_{0}=\int dx\sum_{\alpha\sigma}[\partial_x\Phi(x)]^T\mathbf{M}\partial_x\Phi(x).
\end{equation}
The matrix $\mathbf{M}$ is
\begin{equation}
    \mathbf{M}=\frac{1}{2}
        \begin{pmatrix}
          v_+ & \frac{g_{4+}}{2\pi} & \frac{g_{2\parallel}}{2\pi} & \frac{g_{2\perp}}{2\pi} \\
          \frac{g_{4+}}{2\pi} & v_+ & \frac{g_{2\perp}}{2\pi}& \frac{g_{2\parallel}}{2\pi} \\
          \frac{g_{2\parallel}}{2\pi} & \frac{g_{2\perp}}{2\pi} & v_- & \frac{g_{4-}}{2\pi} \\
          \frac{g_{2\perp}}{2\pi} & \frac{g_{2\parallel}}{2\pi} & \frac{g_{4-}}{2\pi}  & v_-
     \end{pmatrix}\,.
\end{equation}
We will now proceed to diagonalize this step by step.

We begin in the standard way by rotating $\phi_{\alpha\sigma}=(\phi_\sigma-\alpha\theta_\sigma)/\sqrt{2}$ and then $\phi_{c,s}=(\phi_\uparrow\pm\phi_\downarrow)/\sqrt{2}$, and similarly for the $\theta_\sigma$ fields. These of course obey
\begin{equation}
    \left[\phi_{m}(x),\partial_y\theta_{m'}(y)\right]=i\delta_{mm'}\delta(x-y)\,.
\end{equation}
Unlike for a standard Luttinger liquid this does not fully diagonalize the problem as the tilt asymmetry creates a coupling between the field and its conjugate. The diagonal term is, with $H_0=H_{0d}+H_{0t}$,
\begin{equation}
    H_{0d}=\sum_{m\in\{c,s\}}\frac{v_m}{2}\int dx\left[K_m(\partial_x\theta_m)^2+\frac{1}{K_m}(\partial_x\phi_m)^2\right]\,.
\end{equation}
Full expressions for the coefficients can be found in the appendix \ref{app_details}. The asymmetric velocities cause the coupling
\begin{equation}\label{eqtilt}
    H_{0t}=\sum_{m\in\{c,s\}}v_{am}\int dx\partial_x\phi_m(x)\partial_x\theta_m(x)
\end{equation}
where to lowest order
\begin{equation}
    v_{a;c,s}=\frac{v_--v_+}{2}\pm\frac{g_{4-}-g_{4+}}{4\pi}\,.
\end{equation}
Note there can be a different anomalous velocity for the spin and charge sectors but spin and charge remain good quantum numbers for the model.

Finally we must consider the non-oscillatory massive terms. These become
\begin{equation}\label{massive}
    H_M=\frac{g'_{2}}{2\pi^2a^2}\int dx\cos\left[\sqrt{8\pi}\theta_s(x)\right]\,.
\end{equation}
In Sec.~\ref{sec_low} we will analyze this term under renormalization group flow. The total Hamiltonian is $H=H_0+H_M=H_{od}+H_{0t}+H_M$.

To help understand the effect of the tilt $H_{0t}$, we can examine the imaginary-time Green's functions. We consider in particular
\begin{equation}
G_m(x,\tau)=\langle \theta_m(0,0)\theta_m(x,\tau)\rangle
\qquad m\in\{c,s\}.
\end{equation}
as these correlators will also be needed later in the renormalization group analysis. The $\langle\phi_m(0,0)\phi_m(x,\tau)\rangle$ correlator has a related form. Since the theory remains quadratic, the calculation of $G_m(x,\tau)$ is still straightforward. First one can integrate out the field $\phi_m$, followed by a standard Gaussian integral. We find in momentum-frequency space
\begin{equation}
G_m(k,i\omega_n)=\frac{2v_m}{K_m}\,
\frac{1}{\omega_n^2+2i v_{am}k\omega_n+k^2\left(v_m^2-v_{am}^2\right)}.
\end{equation}
The tilt produces the mixed term $2i v_{am}k\omega_n$ and rescales the $k^2$ contribution.

Fourier transforming back to position and imaginary time $\tau$ results in, with $T$ being temperature,
\begin{equation}
G_m(x,\tau)=-\frac{2}{K_m}\ln\left|
\sinh\!\left[\pi T\left(
\frac{v_m|x|}{v_m^2-v_{am}^2}-i\tau
\right)\right]\right|\,.
\end{equation}
We find the standard correlator simply with a rescaled velocity.

The quadratic bosonic Hamiltonian can be fully diagonalised by transforming to the chiral bosonic modes
\begin{equation}
    \phi_{m}=\frac{1}{\sqrt{2K_{m}}}(\varphi_{mR}+\varphi_{mL})
\end{equation}
and
\begin{equation}
    \theta_{m}=\sqrt{\frac{K_{m}}{2}}(-\varphi_{mR}+\varphi_{mL})\,.
\end{equation}
Note that right and left moving chiral fields in the same charge/spin sector do not commute with each other, see appendix \ref{app_details}. We then find
\begin{equation}
    H_0=\sum_{m\in\{c,s\}}\sum_{\delta\in\{R,L\}}\frac{v_{m\delta}}{2}\int dx(\partial_x\varphi_{m\delta})^2\,.
\end{equation}
We have $v_{mR}=v_m-v_{am}$ and $v_{mL}=v_m+v_{am}$. The positive real time Green's functions of the chiral fields, which we will require later for the spectral function, are
\begin{align}\label{chiralg}
    G_{m\delta}(x,t)&=\left\langle\left[\varphi_{m\delta}(x,t)-\varphi_{m\delta}(0,0)\right]^2\right\rangle\\\nonumber&
    =\frac{1}{\pi}\ln\left|\frac{a+i(v_{m\delta}t-\delta x)}{a}\right|\,.
\end{align}
In this form it is clear that the tilt leads to an asymmetry in the velocities of the chiral fields. This asymmetry can be the result of an asymmetry in either the free dispersion or the interaction. In the more standard Luttinger liquid form it manifests as a coupling between the $\phi$ and $\theta$ fields. However as only the velocities are altered, but not the Luttinger parameter no signature of the asymmetry will be found in the scaling of the density of states or impurity backscattering terms.

Finally we can check that our bosonic Hamiltonian has the appropriate symmetry properties. Parity, understood here as spatial inversion $x\to -x$, acts on the chiral bosons as
\begin{equation}
\phi_{\alpha\sigma}\to -\phi_{\bar\alpha\sigma}-\sqrt{\frac{\pi}{8}} \, ,
\end{equation}
which in the charge and spin basis becomes
\begin{equation}
\phi_m\to -\phi_m-\sqrt{\pi},
\qquad
\theta_m\to \theta_m \, .
\end{equation}
Time reversal acts as
\begin{equation}
\phi_{\alpha\sigma}\to -\phi_{\bar\alpha\bar\sigma}
+\sqrt{\frac{\pi}{8}}\,\delta_{\sigma,+} \, ,
\end{equation}
which gives
\begin{align}
\phi_c &\to -\phi_c+\sqrt{\pi},
\qquad
\theta_c \to \theta_c ,\\
\phi_s &\to \phi_s+\sqrt{\pi},
\qquad
\theta_s \to -\theta_s .
\end{align}
Under the combined $PT$ transformation one finds
\begin{align}
\phi_c &\to \phi_c,
\qquad
\theta_c \to \theta_c ,\\
\phi_s &\to -\phi_s,
\qquad
\theta_s \to -\theta_s .
\end{align}
We therefore see that the tilt term $H_{0t}$ breaks parity and time-reversal symmetry separately, while preserving the combined $PT$ symmetry, in agreement with the microscopic motivation of the model.

The bosonized theory also retains the compactification symmetry
\begin{equation}
\phi_{\alpha\sigma}\to \phi_{\alpha\sigma}+\sqrt{2\pi}\,n,
\qquad n\in\mathbb{Z},
\end{equation}
as well as the global $U(1)$ symmetry
\begin{equation}
\phi_{\alpha\sigma}\to \phi_{\alpha\sigma}+\alpha C,
\end{equation}
where $C$ is a constant.

We note that relaxing the $PT$ symmetry requirement will lead to a model where spin and charge are no longer the diagonal modes~\cite{Pereira2004,Sedlmayr2013b}. In general in such a case the tilt term and the usual quadratic modes will likely not be the same. As such the final diagonal chiral modes of the model will be more complicated. However we do not expect any fundamentally new physics to arise in this situation.

\section{Low Energy Hamiltonian}\label{sec_low}

To determine the effective low energy theory for our model we can check the scaling of the $H_M$ term under renormalization group flow. We start from the partition function in a coherent state functional integral form~\cite{Negele1998}:
\begin{equation}
\mathcal{Z}=\int D\phi D\varPi e^{-S}
\end{equation}
where the action is
\begin{equation}
S=\int_0^{\frac{1}{T}} d\tau\left[-i\sum_m\int dx\varPi_m(x)\partial_\tau\phi_m(x)+H\right]
\end{equation}
and $\varPi(x)=\partial_x\theta_m(x)$. Expanding the exponent in terms of $H_M$ and integrating out the charge sector we have
\begin{eqnarray}
\frac{\mathcal{Z}}{\mathcal{Z}_s}\approx1-\frac{g'_2}{2\pi^2a^2}\textrm{Re}\int d\tau dx\left\langle e^{i\sqrt{8\pi}\theta_s} \right\rangle_s\,.
\end{eqnarray}
$\langle\ldots\rangle_s$ refers to the average over the spin action
\begin{eqnarray}
\langle\ldots\rangle_s=\frac{1}{\mathcal{Z}_s}\int D\phi_s D\varPi_s \ldots e^{-S_s}
\end{eqnarray}
where
\begin{align}
S_s=&\int_0^{\frac{1}{T}} d\tau\bigg[-i\int dx\varPi_s(x)\partial_\tau\phi_s(x)+H_s\bigg]\,,\\\nonumber
\mathcal{Z}_s=&\int D\phi_s D\varPi_s e^{-S_s}\,,
\end{align}
and $H_s$ is just the spin sector of $H$.

Following a standard renormalization group procedure let us split the fields into fast, $\theta_s^>$, and slow, $\theta_s^<$, fields. Our fast fields are defined for $\Lambda'<|k|,|\omega|/u<\Lambda$, and the slow for $|k|,|\omega|/u<\Lambda'$ where $\Lambda$ is a cut-off. We integrate over the fast fields and re-exponentiate. Then parameterizing $\Lambda=e^{-l}$ and $\Lambda'=e^{-l-dl}$ the flow equation for $g'_2$ is
\begin{eqnarray}
    \frac{dg'_2}{dl}=g'_2\left[2-\frac{2}{K_s}\right]\,.
\end{eqnarray}
As we have no $SU(2)$ symmetry $K_s>1$ for repulsive interactions and this term is relevant. This is a relevant perturbation and leads to the spin degree of freedom becoming gapped at low energies, bringing us to an effective single mode Luttinger liquid. At low temperatures we can therefore replace the full model with an effective spinless model for the charge sector, with the spin degree of freedom frozen out.

\section{Spectral Function}\label{sec_spec}

A direct probe of the tilt is provided by the momentum-resolved spectral function~\cite{Meden1992,Voit1995}. The spectral function resolves the two chiral edges separately and therefore retains a direct signature of the velocity asymmetry $v_R \neq v_L$. The spectral function is defined as
\begin{equation}
A(k,\omega)=-\frac{1}{\pi}\,\mathrm{Im}\,G^{R}(k,\omega),\,
\end{equation}
in terms of the retarded Green's function,
\begin{equation}
G^{R}(x,t)=\sum_{\alpha\sigma}e^{i\alpha k_Fx}G^{R}_{\alpha\sigma}(x,t)\,,
\end{equation}
which has been resolved into contributions from each spin and Fermi point. For each branch we have
\begin{equation}
G^{R}_{\alpha\sigma}(x,t)=i\left\langle\{\psi_{\alpha\sigma}^\dagger(0,0),\psi_{\alpha\sigma}(x,t)\}\right\rangle\,.
\end{equation}

Following Ref.~\onlinecite{Voit1995} we define
\begin{equation}
    \tilde{G}_{\alpha\sigma}(x,t)=\left\langle\psi_{\alpha\sigma}^\dagger(0,0)\psi_{\alpha\sigma}(x,t)\right\rangle
\end{equation}
so that $G^{R}_{\alpha\sigma}(x,t)=i[\tilde{G}_{\alpha\sigma}(x,t)+\tilde{G}_{\alpha\sigma}(-x,-t)]$. After substituting the vertex operator \eqref{vertex} we can factorize the expressions for the spin and charge sectors:
\begin{align}
    \tilde{G}_{\alpha\sigma}(x,t)=&\frac{1}{2\pi a}\left\langle 
    e^{-i\sqrt{\pi/2}(\alpha\phi_c(0)-\theta_c(0))}
    \right.\\\nonumber&\times\left.
    e^{i\sqrt{\pi/2}(\alpha\phi_c(x,t)-\theta_c(x,t))}\right\rangle_c\\\nonumber&\times
    \left\langle 
    e^{-i\sigma\sqrt{\pi/2}(\alpha\phi_s(0)-\theta_s(0))}
    \right.\\\nonumber&\times\left.
    e^{i\sigma\sqrt{\pi/2}(\alpha\phi_s(x,t)-\theta_s(x,t))}\right\rangle_s\,.
\end{align}
The spin sector is gapped and therefore we focus on the charge sector. We have, rotating to the chiral modes,
\begin{align}
    \tilde{G}_{\alpha\sigma}(x,t)\sim&\left\langle e^{-i\sqrt{\pi}(\xi_{\alpha R}\varphi_{c,R}(0)+\xi_{\alpha L}\varphi_{c,L}(0)}\right.\\\nonumber&\times\left.
    e^{i\sqrt{\pi}(\xi_{\alpha R}\varphi_{c,R}(x,t)+\xi_{\alpha L}\varphi_{c,L}(x,t))} \right\rangle_c\,.
\end{align}
Using the fact that the commutation relations of these field operators are not operators we have
\begin{equation}
    \tilde{G}_{\alpha\sigma}(x,t)\sim 
    \left\langle e^{i\sqrt{\pi}\sum_\delta\sqrt{\xi_{\alpha \delta}}[\varphi_{c\delta}(x,t)-\varphi_{c\delta}(0,0)]} \right\rangle_c
\end{equation}
where
\begin{eqnarray}
    \sqrt{\xi_{\alpha\delta}}=\frac{1}{2}\left(\frac{\alpha}{\sqrt{K_c}}+\delta\sqrt{K_c}\right)\,.
\end{eqnarray}
Therefore we finally find
\begin{equation}\label{gt}
    \tilde{G}_{\alpha\sigma}(x,t)\sim e^{-\pi\sum_\delta\xi_{\alpha \delta}\left\langle[\varphi_{c\delta}(x,t)-\varphi_{c\delta}(0,0)]^2 \right\rangle_c}
\end{equation}
for these Green's functions.

\begin{figure}
    \centering
\includegraphics[width=0.9\columnwidth]{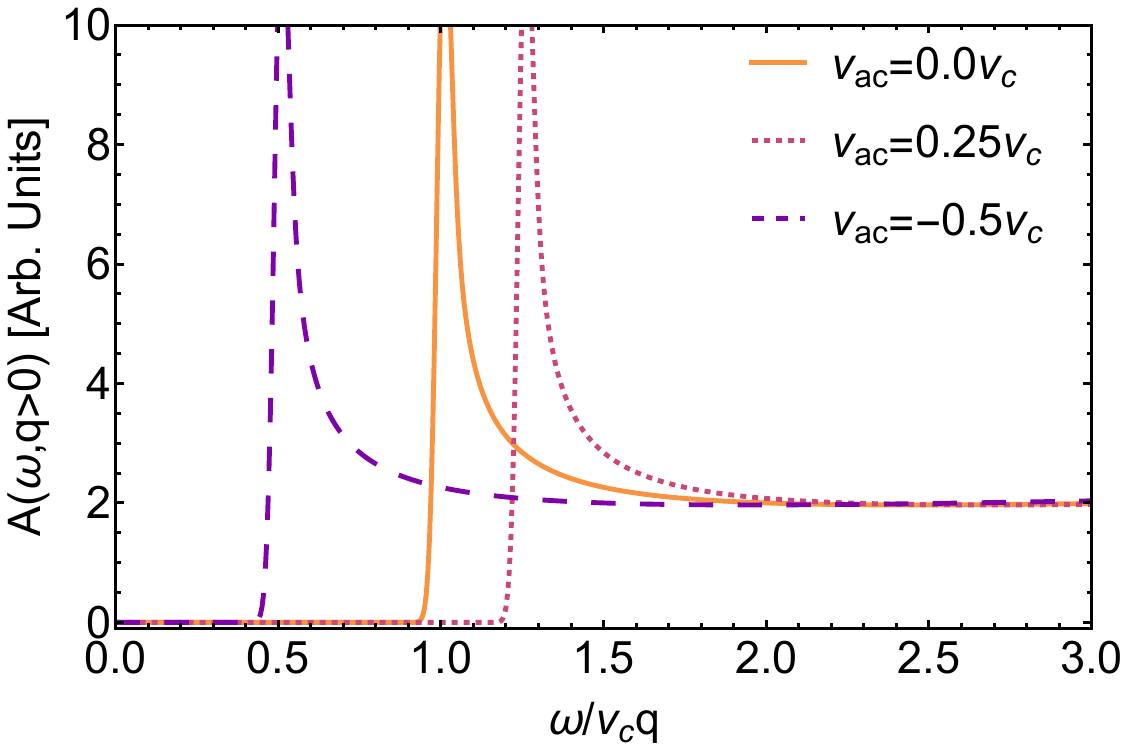}\\
\includegraphics[width=0.9\columnwidth]{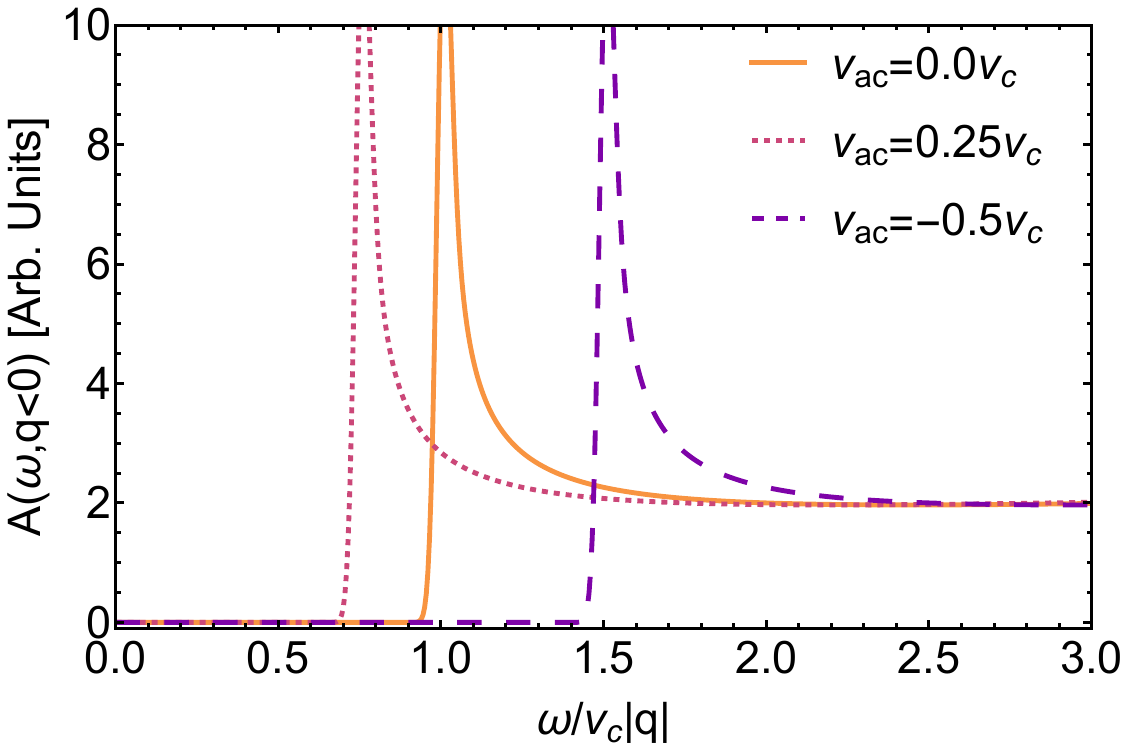}
\caption{The spectral functions for $q>0$, top panel, and $q<0$, bottom panel. Relatively large values of the asymmetric velocity $v_{ac}$ are used for illustrative purposes. For plotting the spectral function has been convoluted with a Gaussian resolution function of width $\omega/v_c|q| = 0.02$.}\label{fig_spec}
\end{figure}

Using the chiral bosonic Green's functions \eqref{chiralg} and \eqref{gt} the spectral function becomes
\begin{equation}
    A(k,\omega)\sim\textrm{Re}\sum_{\alpha}\int dx dt e^{i\omega t-i(k-\alpha k_F)x}\prod_\delta\left|\frac{a}{x-\delta v_{c\delta}t}\right|^{\xi_{\alpha\delta}}
\end{equation}
Resolving near the two Fermi points we can define the contributions $A_+(q=k-k_F,\omega)$ and $A_-(q=k+k_F,\omega)$. Up to a non-universal cutoff-dependent prefactor one finds for the singular contributions
\begin{equation}
A(q>0,\omega)\sim
\Theta(\omega-v_{cR} q)\sum_\alpha\left|\omega-v_{cR} q\right|^{\xi_{\alpha R}-1}\,,
\label{Ap_final}
\end{equation}
and
\begin{equation}
A(q<0,\omega)\sim
\Theta(\omega+v_{cL})q)\sum_\alpha\left|\omega+v_{cL}q\right|^{\xi_{\alpha L}-1}\,.
\label{An_final}
\end{equation}
We note that as required in the limit that there is no torsion the spectral function thus recovers the usual singular dependency. In Fig.~\ref{fig_spec} examples of the spectral function are shown for different asymmetries. By comparison of the positive and negative momentum dependence a signature of the tilt can be found.

\begin{figure}
    \centering
\includegraphics[width=0.9\columnwidth]{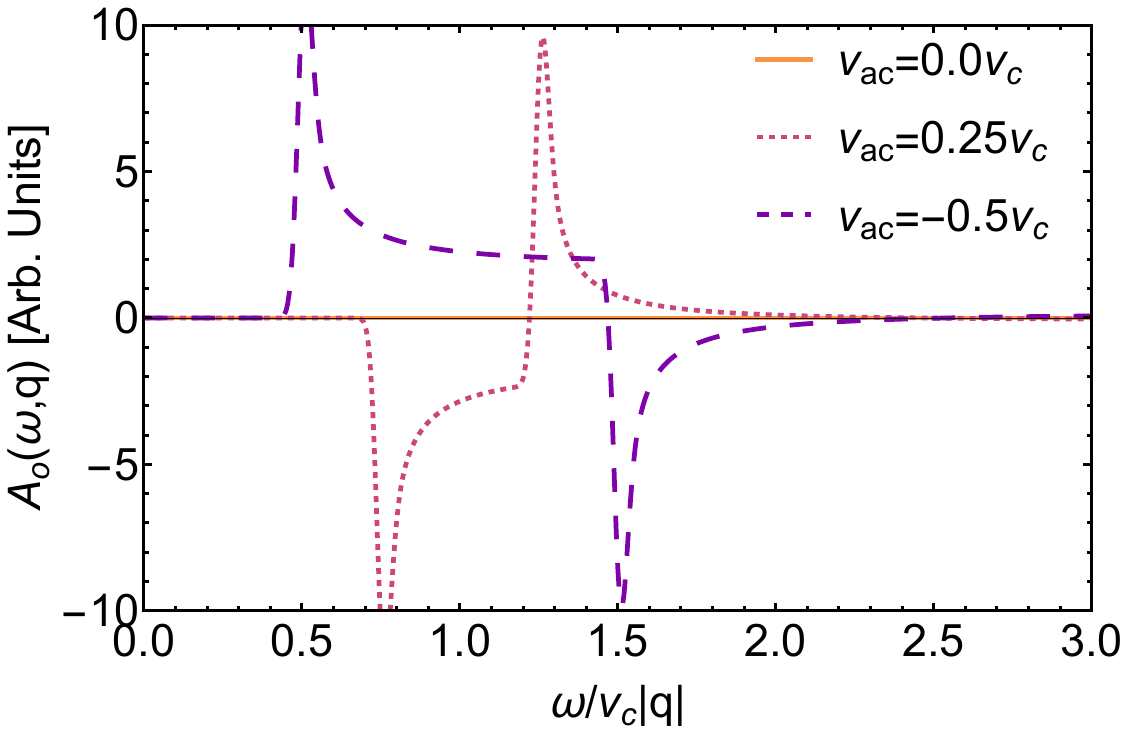}
\caption{The odd spectral function, see Eq.~\eqref{s_e_odd}. A non-zero odd spectral function is a demonstration of the tilt. For plotting the spectral function has been convoluted with a Gaussian resolution function of width $\omega/v_c|q| = 0.02$. See teh caption of Fig.~\ref{fig_spec} for more details.}\label{fig_spec_eo}
\end{figure}

As the signal of the tilt is an asymmetry in the velocities of the left and right moving chiral bosonic modes, it makes sense to define even and odd combinations:
\begin{equation}
    A_{e}(\omega,q)=\frac{A(q,\omega)+A(-q,\omega)}{2}
\end{equation}
and
\begin{equation}\label{s_e_odd}
    A_{o}(\omega,q)= A(q,\omega)-A(-q,\omega)\,.
\end{equation}
In the absence of torsion We should find that $A_e(\omega,q)=A(\omega,q)$ and $A_o(\omega,q)=0$. Thus $A_o(\omega,q)\neq0$ is a signature of the tilt, see Fig.~\ref{fig_spec_eo} for an illustration.

\section{Conclusions and Discussion}\label{sec_con}

In this article we have introduced a general model of spinfull interacting electrons with an asymmetric dispersion which breaks parity and time-reversal symmetries but not their composite. We have given a microscopic derivation for how such a model can be achieved from a helical wire with torsion and Zeeman splitting, and derived the appropriate Luttinger liquid model. Including all general interacting terms we show via a first order renormalization group analysis that generically the spin sector is gapped at low energies, resulting in an effective spinless Luttinger liquid ground state.

The Luttinger liquid is characterized by asymmetric velocities for the left and right moving chiral bosonic modes. These modes remain diagonal. In the typical language of the fields and conjugate fields the asymmetry manifests as a coupling term proportional to the velocity asymmetry. The Luttinger parameters however remain unaffected, which means that no signature of the asymmetry can easily seen in impurity scattering, differential conductance, or the density of states. Though prefactors of these terms may depend on the velocity asymmetry, the prefactors in any case depend on cut-offs and can not be used to determine the presence of the asymmetry. Instead we demonstrate that the spectral function contains signatures of the asymmetry which would in principle be accessible experimentally.

\appendix

\section{Bosonization Details}\label{app_details}

In this appendix we summarize and also include some further details of the models and bosonization procedure. The vertex operator is
\begin{equation}
    \psi_{\alpha\sigma}(x)=\frac{1}{\sqrt{2\pi a}}e^{i\alpha\sqrt{2\pi}\phi_{\alpha\sigma}(x)}
\end{equation}
and the field $\phi(x)$ obeys
\begin{equation}
    \left[\phi_{\alpha\sigma}(x),\phi_{\alpha'\sigma'}(x')\right]=\delta_{\alpha\alpha'}\delta_{\sigma\sigma'}\frac{i\alpha}{2}\sgn(x-x')\,.
\end{equation}
Additionally
\begin{equation}
    \sqrt{2\pi}\rho_{\alpha\sigma}=-\partial_x\phi_{\alpha\sigma}
\end{equation}
for the fermionic density.

Following bosonization using the vertex operator a series of rotations are made to simplify the quadratic bosonic Hamiltonian. Compiling these operations we find
\begin{equation}
    \phi_{\alpha\sigma}=\frac{\phi_c+\sigma\phi_s-\alpha(\theta_c+\sigma\theta_s)}{2}\,,
\end{equation}
for the relation between the original bosonic field and the conjugate fields for spin and charge. The fields obey
\begin{equation}
    \left[\phi_{m}(x),\partial_y\theta_{m'}(y)\right]=i\delta_{mm'}\delta(x-y)\,.
\end{equation}
The asymmetric Hamiltonian can be fully diagonalized by rotating to the ``chiral'' fields $\varphi_{m\delta}$ via
\begin{equation}
    \phi_{m}=\frac{1}{\sqrt{2K_{m}}}(\varphi_{mR}+\varphi_{mL})
\end{equation}
and
\begin{equation}
    \theta_{m}=\sqrt{\frac{K_{m}}{2}}(-\varphi_{mR}+\varphi_{mL})\,.
\end{equation}
The inverse relation is given by
\begin{equation}
    \varphi_{m\delta}=\frac{1}{\sqrt2}\left(\sqrt{K_m}\phi_m-\delta\frac{\theta_m}{\sqrt{K_m}}\right)
\end{equation}
and the commutation relation is
\begin{equation}
    \left[\varphi_{m\delta}(x),\varphi_{m'\delta'}(x')\right]=\frac{\delta i}{2}\delta_{mm'}\delta_{\delta\delta'}\sgn(x-x')\,.
\end{equation}
Here, as elsewhere $m\in\{c,s\}$ labels the spin and charge modes and $\delta\in\{L,R\}$ labels the left and right moving chiral \emph{bosonic} fields. Where convenient $\delta$ should be understood to have the values $\pm1$ for R/L movers in mathematical expressions. $\alpha\in\{+,-\}$ labels the Fermi points of the non-interacting theory.

It is convenient to define
\begin{equation}
\bar v \equiv \frac{v_++v_-}{2}\,
\textrm{, and }
\delta v \equiv \frac{v_- - v_+}{2}.
\end{equation}
Similarly:
\begin{align}
g_{4\Sigma}&\equiv g_{4+}+g_{4-}\,,\\\nonumber
\delta g_4&\equiv g_{4-}-g_{4+}\,,\\\nonumber
g_{2c}&\equiv g_{2\parallel}+g_{2\perp}\,\textrm{, and}\\\nonumber
g_{2s}&\equiv g_{2\parallel}-g_{2\perp}\,.
\end{align}
Then the velocities and Luttinger parameters are
\begin{align}
v_c&=
\sqrt{
\left(\bar v+\frac{g_{4\Sigma}}{4\pi}\right)^2
-
\left(\frac{g_{2c}}{2\pi}\right)^2
}\,,\\\nonumber
K_c&=\sqrt{\frac{4\pi\bar v+g_{4\Sigma}-2g_{2c}}
{4\pi\bar v+g_{4\Sigma}+2g_{2c}}}\,,\\\nonumber
v_s&=\sqrt{\left(\bar v-\frac{g_{4\Sigma}}{4\pi}\right)^2
-\left(\frac{g_{2s}}{2\pi}\right)^2}\,,\textrm{ and}\\\nonumber
K_s&=\sqrt{\frac{4\pi\bar v-g_{4\Sigma}-2g_{2s}}
{4\pi\bar v-g_{4\Sigma}+2g_{2s}}}\,.
\label{app:Ks}
\end{align}
The asymmetry is encoded in the velocities
\begin{equation}
v_{ac}=\delta v+\frac{\delta g_4}{4\pi}=
\frac{v_- - v_+}{2}+\frac{g_{4-}-g_{4+}}{4\pi}
\label{app:vac}
\end{equation}
and
\begin{equation}
v_{as}=\delta v-\frac{\delta g_4}{4\pi}=
\frac{v_- - v_+}{2}-\frac{g_{4-}-g_{4+}}{4\pi}
\label{app:vas}
\end{equation}
which contribute to $H_{0t}$, Eq.~\eqref{eqtilt}.


%

\end{document}